# Tunable Low Density Palladium Nanowire Foams

*Dustin A. Gilbert,[†‡] Edward C. Burks,[†] Sergey V. Ushakov,[§] Patricia Abellan,[||] Ilke Arslan,[⊥] Thomas E. Felter,[∇] Alexandra Navrotsky,[§] and Kai Liu[†]\**

[†] Department of Physics, University of California, Davis, CA 95616

[‡] NIST Center for Neutron Research, National Institute of Standards and Technology, Gaithersburg MD 20899

[§] Peter A. Rock Thermochemistry Laboratory and NEAT ORU, University of California, Davis CA 95616

[||] SuperSTEM Laboratory, SciTech Daresbury Campus, Daresbury, U.K.

[⊥] Physical Sciences Division, Pacific Northwest National Laboratory, Richland, WA 99352

[∇] Energy Innovation Department 8366, Sandia National Laboratory, Livermore CA 94550

ABSTRACT

Nanostructured palladium foams offer exciting potential for applications in diverse fields such as catalyst, fuel cell, and particularly hydrogen storage technologies. We have fabricated palladium nanowire foams using a cross-linking and freeze-drying technique. These foams have a tunable density down to 0.1% of the bulk, and a surface area to volume ratio of up to $1.54 \times 10^6$:1. They exhibit highly attractive characteristics for hydrogen storage, in terms of loading capacity, rate of absorption and heat of absorption. The hydrogen absorption/desorption process is hysteretic in nature, accompanied by substantial lattice expansion/contraction as the foam converts between Pd and PdH$_x$.

Low density (LD) materials present exciting opportunities in diverse fields such as catalysts,[1] electronics,[2-3] energy production and storage,[4-6] heat exchange,[7] and structural materials.[8-9] Metal and metal oxide foams are particularly appealing due to their high thermal and electrical conductivity, structural strength, environmental and thermal stability, and specific chemical properties. Palladium foams, for example, are suitable for organic chemical synthesis,[10-13] pollution control,[14-15] hydrogen storage,[6, 16] fuel cells,[4] and batteries,[17-18] where a high surface area (SA) to volume ratio (SA:V) is desirable. However, LD metal foams fabricated by traditional techniques are prone to contamination[19-20] or poor crystallinity,[21] negatively affecting their functionality.[22-23] Furthermore, it is difficult to achieve LD metal foams with tunable densities which is crucial for their incorporation into a broad range of technologies.

Recently, LD nanowire foams were realized using a classical sol-gel technique applied to nanowire suspensions.[24-25] Their densities can be as low as a few percent of the bulk density[24] and are determined by the gelling density of the suspension, which is not easily tunable. Furthermore, this approach requires moderately high temperature and pressure processing, and prolonged immersion in organic solvents. A recent work has improved upon this approach by using a freeze-cast technique to fabricate Cu nanowire monoliths.[26]

Here we report the synthesis of LD palladium nanowire foams with density down to 0.1% of the bulk, fabricated by a freeze-casting technique. This synthesis approach does not use any caustic chemicals or high temperature/pressure treatments, and can be applied generally to any aqueous dispersion of nanowires. Palladium foam monoliths demonstrate excellent characteristics for hydrogen storage applications, including their hydrogen loading capacity, rate of absorption, and heat of absorption. The hydrogen absorption/desorption process is highly hysteretic, along with substantial lattice expansion/contraction as the foam converts between Pd and $PdH_x$.

Polycrystalline Pd nanowires were fabricated by electrodeposition into porous templates, then harvested from the templates and suspended in water [Figure 1(a-i)] as discussed in the Experiment section. Transmission electron microscopy (Supporting Information) shows that the Pd crystallites possess a preferred (111) orientation along the wire growth direction. To construct the nanowire foam, the wires were allowed to precipitate out of solution [Figure 1(a-ii)] and the water level was adjusted to the nominal final volume of the foam [Figure 1(a-iii)]. The settled nanowires were then sonicated to develop a randomly dispersed slurry [Figure 1(a-iv)], which was immediately immersed in liquid nitrogen, freezing the wires in random distribution in the ice matrix [Figure 1(a-v)]. The frozen slurry was then placed in vacuum (< 0.1 Pa) for >12 hours to sublimate the interstitial ice, leading to the pure Pd nanowire foam [Figure 1(a-vi)]. SEM micrographs [Figure 1(b, c)] confirm the highly porous structure of the fabricated foams. For Pd foams comprised of 200 nm diameter, 15 μm long wires, the foam density was tunable between (12-135) mg cm$^{-3}$, corresponding to a density that is only 0.1% - 1% of bulk Pd. This synthesis approach is scalable for industrial applications, as wet chemistry based solution synthesis is well suited for mass production of nanowires and industrial size freeze-drying apparatus is readily available. In this work we focus on an example monolith which has a density of 41±3 mg cm$^{-3}$ and SA of 6.9±0.1 m$^2$ g (Pd)$^{-1}$ (the confidence interval represents one standard deviation, and is determined by the accuracy of the measurement tools); porosity measurements show that the foam has few constricted volumes (e.g. cavities or cracks, further details discussed in Supporting Information).

The X-ray diffraction (XRD) pattern of the Pd nanowires [Figure 2(a)] shows the Pd (111) and (200) peaks, identifying a cubic lattice parameter of $a$ = 3.86 ± 0.01 Å, consistent with bulk values.[27] After a 1 hour exposure to ≈200 kPa hydrogen (at 25 °C) the peaks shift to lower $2\theta$ values as $a$ increases to 4.00 ± 0.01 Å, signaling the formation of PdH.[27] Compared to thermal expansion, this lattice expansion is quite drastic and may produce large stress within the wires. The hydrogen can be desorbed by heating (250 °C for 30 min in air) or vacuum (30 min, 25 °C, $P_{Base}$< 0.01 Pa), as indicated by the shifting XRD peaks. Translation of the peaks, rather than broadening or splitting, indicates that the hydrogen penetrates the entire wire uniformly, compared to bulk palladium in which penetration is limited by the rate of hydrogen diffusion. Interestingly, after successive hydrogen exposures, the PdH (111) peak shifts to lower angles, indicating the lattice parameter for the PdH increases with cycling, corresponding to enhanced hydrogen absorption.

After several hydrogenation/dehydrogenation cycles on the monolith, the SA was found to have increased to 11.4 ± 0.5 m$^2$ g$^{-1}$ (4.67 × 10$^5$ m$^{-1}$), a change of 65 %. SEM micrographs [Figure 2(b)] reveal that fractures have developed along the nanowires, likely caused by strain due to the substantial lattice expansion during hydrogenation, as shown by XRD. Similar cracking has been reported in palladium filter membranes during hydrogen cycling.[28] While the fracturing is an unexpected effect, the enhanced SA:V may be beneficial for some applications which rely on large SA. For such fractured nanowires, the maximal SA:V of the foam is increased to $(1.54\pm0.07)\times10^6$ m$^{-1}$ (for a 135 mg cm$^{-3}$ foam at 11.4 m$^2$ g$^{-1}$).

Hydrogen absorption/desorption isotherms measured on the aforementioned 41 mg cm$^{-3}$ palladium foam are shown in Figure 2(c). Hydrogen absorption is shown to begin at ≈1.6 kPa, and continue until ≈2.1 kPa.[27] The hydrogen storage capacity at normal temperature and pressure (NTP) was measured to be 2.88 ± 0.07 mmol(H$_2$) g(Pd)$^{-1}$, a H:Pd ratio of 0.61. These values are comparable to literature values [3.4 mmol(H$_2$) g(Pd)$^{-1}$, H:Pd ratio of 0.72], demonstrated in bulk Pd.[27, 29] Along the descending branch of the isotherm, desorption occurs at 0.73 kPa, showing a hysteretic absorption centered at 1.26 kPa, with a hysteresis of 880 Pa.

For comparison, hydrogen isotherms were also taken on porous palladium micropowders [Figure 2(c)], with the microstructure shown in Figure 2(d). The micropowder isotherm has a wider hysteresis (1.44 kPa) and is centered at a higher pressure (1.38 kPa) compared to the foam. The maximal hydrogen storage capacity at NTP is measured to be 3.06±0.01 mmol(H$_2$) g(Pd)$^{-1}$, a H:Pd ratio of 0.65, also comparable to our nanowire foam.

Rate of absorption (ROA) measurements on the 41 mg cm$^{-3}$ foam [Figure 3] show an asymptotic decay in pressure, which can be modeled as $P(t)=a_1e^{-t/c1}+a_2e^{-t/c2}+b$, where $a_{1,2}$ and $c_{1,2}$ are the relative weight and time constants of each mechanism, and $b$ identifies the static equilibrium pressure. The dynamic pressure, $P_{\text{Measured}}-b$, is shown in Figure 3(a, b) for the foam and powder, respectively, and static equilibrium pressure is shown in Figure 3(c). The equilibrium pressure confirms that the measurements proceed from the un-hydrogenated Pd state (I), through the hydrogen absorption phase (II) and into the α' PdH$_x$ region (III). The decay constants ($c_1$, $c_2$) from the proposed model ($\chi^2$ = 0.98) are shown in Figure 3(d) and identify fast ($c_1 \approx$ 2 s) and slow ($c_2 \approx$ 40 s) dynamic mechanisms. Interestingly, the decay constants are similar for both the foam and powder, eluding to their common physical origin. The relative weight coefficients ($a_1/a_1^{P=0}$, $a_2/a_1^{P=0}$) [Figure 3(e)] identify the relative activity of each of the mechanisms.

In both the foam and powder the initial dosage is dominated by the fast mechanism ($c_1$), which can be attributed to surface adsorption; this mechanism is active throughout the measurements (I-III). However, for the foam the slow mechanism ($c_2$) is active only during hydrogen absorption (region II), as identified by the vanishing $a_2$. In comparison, for the powder the slow mechanism is active (non-zero $a_2$) throughout the isotherm except during the initial exposure. This may suggest a broader distribution of absorption potentials resulting from defect sites or internal strain in the powder.

Calorimetric measurements of heat of incorporation of hydrogen ($\Delta H$) were performed under increasing hydrogen pressure, as shown in Figure 3(f), with regions transposed from the pressure isotherm in Figure 3(c). For both the foam and powder the initial exposure to hydrogen causes an abrupt increase in the magnitude of $\Delta H$, followed by a gradual decrease. In this region $\Delta H$ is associated with the adsorption and dissociation of H$_2$ on the Pd surface and reaches an extremum of -50 ± 0.5 kJ mmol(H$_2$)$^{-1}$ in the foam and -40 ± 0.5 kJ mmol(H$_2$)$^{-1}$ for the powder. In region II $\Delta H$, representing the heat of formation of palladium hydride, $\Delta H_{PdH}$, becomes approximately constant at -34 ± 0.5 kJ/mol(H$_2$) and -36 ± 0.5 kJ mol(H$_2$)$^{-1}$ for the foam and powder, respectively. This value is in good agreement with previously reported values of -38 kJ mol(H$_2$)$^{-1}$.[27]

The integral excess Gibb's free energy, $\Delta \overline{G}_H^E$, is calculated from the absorption isotherms following procedures discussed previously[30-31] (Supporting Information) to be -9.1 ± 0.5 kJ mol(H$_2$)$^{-1}$ and -7.6 ± 0.5 kJ mol(H$_2$)$^{-1}$ for the foam and powder, respectively. These values are slightly less negative than the previously reported value of -13.9 kJ mol(H$_2$)$^{-1}$ obtained from Pd foils.[32] From the isotherm which gives free energy and the measured enthalpy, the excess entropy of formation, $\Delta S_{PdH}$, is calculated to be -51 ± 2 J mol(H$_2$)$^{-1}$ K$^{-1}$ and -67 ± 2 J mol(H$_2$)$^{-1}$ K$^{-1}$ for the foam and powder, respectively, consistent with previous studies.[33] Interestingly $\Delta \overline{G}_H^E$ calculation also suggests a repulsive H-H interaction at $P > 2.5$ kPa (Supporting Information). This is a significant deviation from previous works on bulk materials, which report an attractive H-H interaction[31-32] and may indicate a deviation from the standard hydrogen cluster absorption model.[31-32]

Lastly, we investigate hysteresis in the desorption process using the first order reversal curve (FORC) technique,[34] which can provide insight into intrinsic properties and interactions in hysteretic systems.[35-40] Using a procedure analogous to previous studies, the sample is initially prepared in the un-hydrogenated state, then dosed with hydrogen, driving it to a *reversal pressure, $P_R$,* at which point the sample is in a mixed state of Pd and PdH$_x$. After reaching $P_R$, the *applied pressure, $P$,* is decreased while measuring the net absorption, $Q(P, P_R)$, and the sample is brought back to the un-hydrogenated state, thus tracing out a single FORC. A family of FORCs is measured by repeating this process for successively higher $P_R$ until the saturated single-phase α'-PdH state is reached. The FORC distribution is extracted by taking the derivative

$$\rho(P, P_R) = \frac{1}{2} \frac{\partial^2 Q(P, P_R)}{\partial P \partial P_R}.$$ The FORC distribution thus maps out the changes of $dQ/dP$ (corresponding to hydrogen desorption) which have a dependence on $P_R$; alternatively, FORC identifies desorption events which are absent on neighboring FORCs. Following this measurement procedure the $P$ axis identifies hydrogen desorption, while the $P_R$ axis identifies hydrogen absorption. By measuring FORCs between the Pd and α' PdH states we gain insight into the intrinsic properties of the system, e.g., due to impurities or crystallographic orientations, and interactions, e.g. properties that depend on the absorption state, including $PdH_x$ lattice strain and H-H interactions.

The family of FORCs of the 41 mg cm$^{-3}$ foam [Figure 4(a)] shows that irreversible absorption begins at 0.93 kPa and subsequent desorption occurs at 0.73 kPa, generating a feature in the FORC distribution at ($P$=0.73 kPa, $P_R$=0.93 kPa) [Figure 4(b)]; the absorption/desorption hysteresis for this initial event is a mere 0.2 kPa, indicating weak binding between the hydrogen and Pd. The first desorption event along each FORC, approximately indicated by the dashed arrow in Figure 4(b), occurs at progressively higher pressures until $P_R >$ 1.4 kPa, at which point the onset of the desorption remains constant at $P =$ 1.1 kPa. This indicates that, for $P_R <$ 1.4 kPa the hydrogen more readily desorbs (i.e. at higher pressures) as the PdH phase fraction increases. This may suggest that both the Pd and the PdH phases could be changing composition, and a combination of thermodynamic and kinetic factors may be at play.

In summary, we have fabricated palladium nanowire foams using a freeze-casting technique with a tunable SA:V of up-to 1.54×10$^6$ m$^{-1}$ and density of 0.1%-1% of the bulk. This technique requires no caustic chemicals or high temperature/pressure processing, and is scalable for industrial applications. These foams demonstrate excellent hydrogen storage characteristics, including loading capacity, rate of absorption, and calorimetry. The hysteretic hydrogen absorption/desorption process is coupled with substantial lattice expansion/contraction as the foam converts between Pd and PdH$_x$. Such foams with pristine metal surfaces are also suitable for catalysis as well as biomedical applications, including cell scaffolding[41-43] and drug delivery.[44]

**Methods**

Palladium nanowires were fabricated by electrodeposition from an aqueous solution of 6 mM PdCl$_2$ + 0.1 M HCl (1 M = 1 mol L$^{-1}$). Electrodeposition was performed at -450 mV relative to a Ag$^+$/AgCl reference electrode into Au-coated (working electrode) anodized aluminum oxide (AAO) or track-etched polycarbonate membranes.[45-46] Nanowires with diameters of (10 - 200) nm and lengths of (3 - 20) μm were achieved. After deposition, the Au working electrode was selectively etched using a solution of 0.4 M K$_3$Fe(CN)$_6$ + 0.2 M KCN +0.1 M KOH. The AAO (polycarbonate) membranes were then dissolved by sonicating them in 6 M NaOH (dichloromethane). The nanowires were transferred to distilled or de-ionized water using a precipitation/decanting/solvent replacement technique. Nanowire foams were then freeze-cast into foam monoliths as discussed in the main text. The mechanical strength of the foam can be further enhanced by sintering,[47] as discussed in the Supporting Information.

X-ray diffraction was performed with Cu K$\alpha$ radiation ($\lambda$=1.54 Å). Scanning electron microscopy (SEM, 5 keV), high resolution transmission electron microscopy (200 kV) was performed on both the foams and individual wires.

Surface area and pore volume distributions were determined from krypton (Kr) gas adsorption isotherms at -196 °C using BET (Branuer-Emmet-Teller)[48] and BJH (Barrett-Joyner-Halenda)[49] methods, respectively. Hydrogen absorption/desorption measurements were performed using a commercially available precision gas dose controller with forked sample tube and a Calvet-type twin microcalorimeter, described previously.[50] Equilibrium was defined as a pressure change of <0.01% over 10 s. Thermodynamic measurements were performed using a constant temperature (37 °C), incremental dosing approach and by integrating the heat flow from the calorimeter.[50] Rate of adsorption (ROA) measurements were performed by exposing the samples to 0.1 mmol($H_2$) g(Pd)$^{-1}$ and recording the pressure for 200 s.

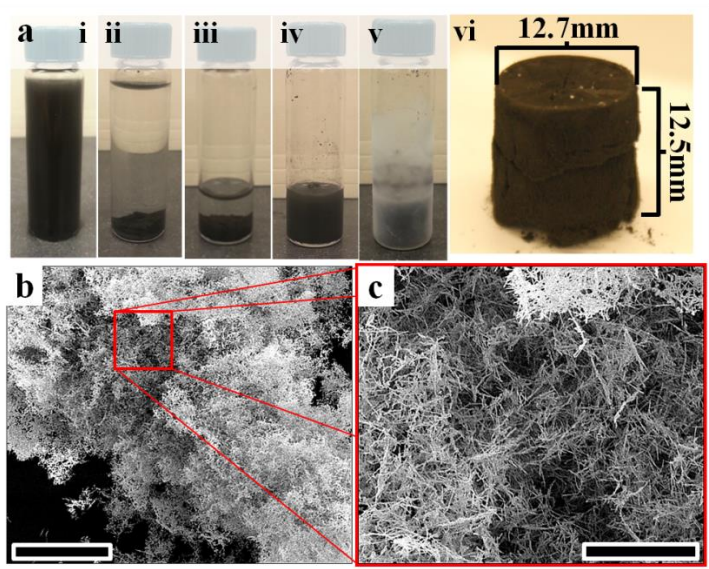

**Figure 1.** (a) Process for fabricating the nanowire foams. (b, c) SEM images of the nanowire foam showing the low-density structure. Scale bar indicates (b) 150 μm and (c) 30 μm.

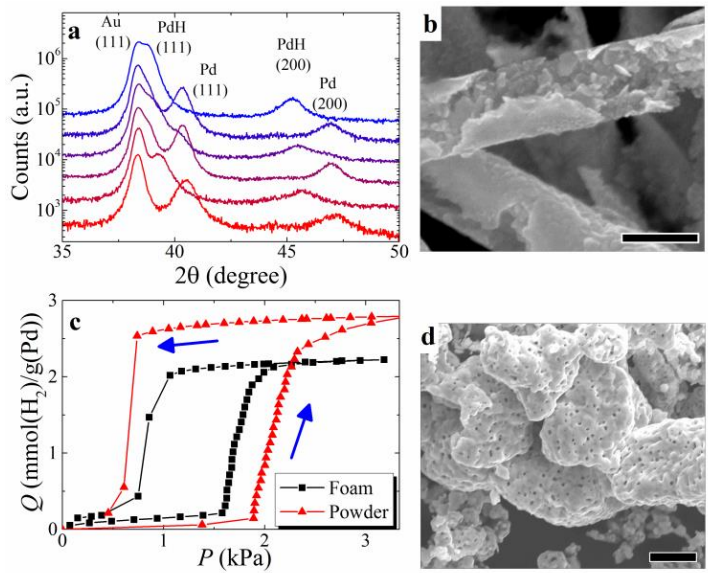

**Figure 2.** (a) X-ray diffraction patterns of Pd nanowires drop-cast on an Au-on-glass substrate showing sequential treatments (red to blue): as-grown, hydrogenation (200 kPa H$_2$, 25 °C, 30 min), heating (250 °C for 30 min), hydrogenation, vacuum (P<0.01 Pa, 25 °C, 30 min), hydrogenation. (b) SEM image of the Pd nanowires after the hydrogenation/dehydrogenation cycles discussed in (a), where the scale bar indicates 200 nm. (c) Hydrogen absorption/desorption isotherm. (d) SEM image of the nanoporous Pd powder with the scale bar indicating 5 μm.

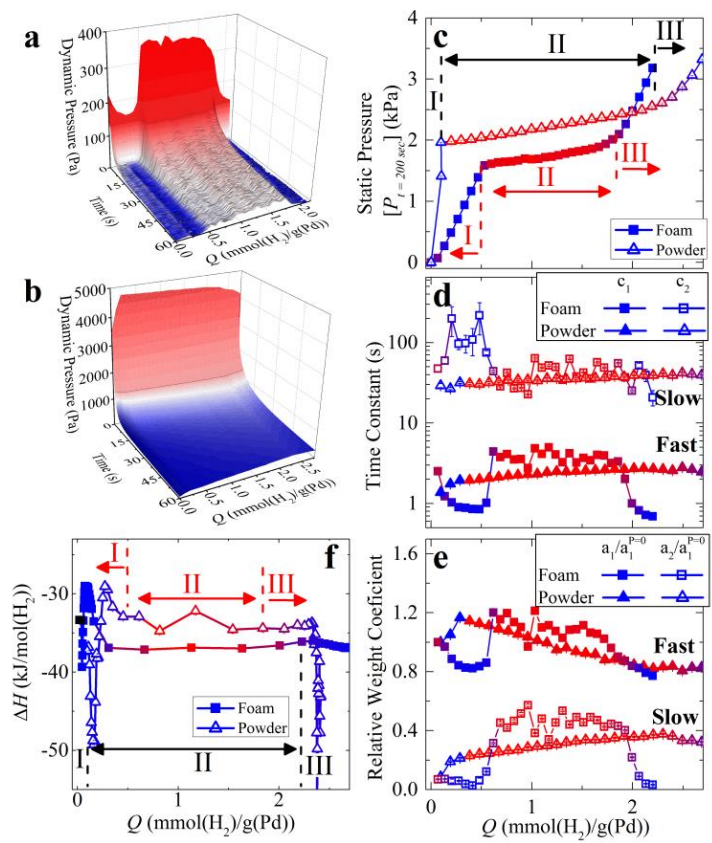

**Figure 3.** Dynamic pressure for (a) Pd foam and (b) Pd powder. Fitted model parameters for (c) equilibrium pressure, (d) time constants and (e) weight coefficient. (f) Calorimetry measurements showing ΔH vs. absorption; region delineators were taken from (c). Color in (a, b) is defined by the spontaneous dynamic pressure P(t); in (c-f) is defined by the derivative of (c) and used to identify the absorption region. Error bars in (c) and (f) are determined by the accuracy of the measurement tools; error bars in (d) and (e) represent one standard deviation from the fitted parameters.

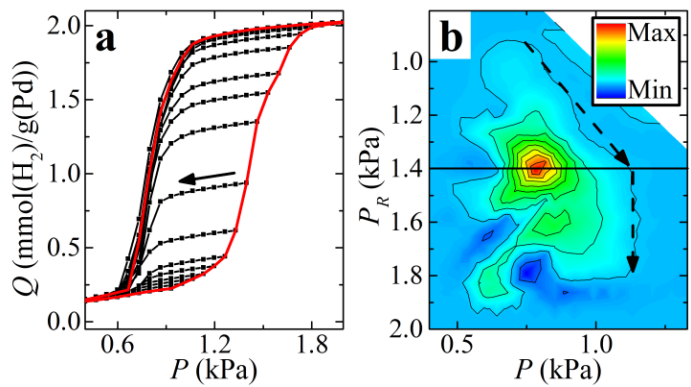

**Figure 4.** (a) Family of FORC's and (b) FORC distribution $\rho(P,P_R)$, mapping out the distribution of absorption/desorption events. Solid arrow in (a) indicates the direction of the FORC measurement at each reversal pressure $P_R$. Dashed arrows in (b) identify the high-pressure boundary of the FORC feature, defined by the weakest (10%) contour line, representing the approximate pressure at which desorption begins.

ASSOCIATED CONTENT

**Supporting Information**.

The following files are available free of charge.

> Figures showing nanostructure of pristine Pd nanowires and sintered films, and Gibbs free energy calculation PDF

AUTHOR INFORMATION

**Corresponding Author**

*Kai Liu kailiu@ucdavis.edu*Kai Liu kailiu@ucdavis.edu

**Author Contributions**

D.A.G. and K.L. conceived and designed the project. Samples were fabricated by D.A.G. and E.C.B. Transmission electron microscopy studies were performed by P.A. and I.A. Structural change under hydrogen absorption/desorption was analyzed by x-ray diffraction and scanning electron microscopy by D.A.G., E.C.B., T.E.F. and K.L. Gas absorption isotherms, surface area and calorimetry studies were performed by S.V.U., D.A.G., and A.N. D.A.G. and K.L. wrote the first draft of the manuscript. All authors contributed to the data analysis and revision of the manuscript. All authors have given approval to the final version of the manuscript.

ACKNOWLEDGMENT

This work was supported by DTRA (Grant #BRCALL08-Per3-C-2-0006, for synthesis), and in part by NSF (DMR-1008791 and DMR-1610060, for characterizations). The calorimetry studies were supported by the Department of Energy, Office of Science, Grant DE-FG02-03ER46053. The Pacific Northwest National Laboratory is operated by Battelle for the US Department of Energy under contract DE-AC05-76RL01830. SuperSTEM is the UK EPSRC National Facility for Aberration-Corrected STEM, supported by the Engineering and Physical Science Research Council.